\documentclass[a4paper,12pt]{article}
\title{Index for $T_{1,1}$ SCFT}
\author{Yu Nakayama}
\usepackage{amsmath}
\usepackage{amssymb}
\usepackage{graphicx}
\def\drawbox#1#2{\hrule height#2pt
        \hbox{\vrule width#2pt height#1pt \kern#1pt
              \vrule width#2pt}
              \hrule height#2pt}

\def\Fund#1#2{\vcenter{\vbox{\drawbox{#1}{#2}}}}
\def\Asym#1#2{\vcenter{\vbox{\drawbox{#1}{#2}
              \kern-#2pt       
              \drawbox{#1}{#2}}}}

\def\funda{\Fund{6.5}{0.4}}

\def\symm{\funda\kern-0.4pt\funda}

\allowdisplaybreaks[0]

\newcommand{\cN}{{\cal N}}

\newcommand{\sectiono}[1]{\section{#1}\setcounter{equation}{0}}

\begin{document}

\begin{titlepage}
\thispagestyle{empty}
\begin{flushright}
UT-06-02\\
hep-th/0602284\\
\end{flushright}

\vskip 1.5 cm

\begin{center}
\noindent{\textbf{\LARGE{\\\vspace{0.5cm} Index for Supergravity on $AdS_5 \times T^{1,1}$
}}} 
\textbf{\LARGE{\\\vspace{0.5cm} and Conifold Gauge Theory
}}
\vskip 1.5cm
\noindent{\large{Yu Nakayama}\footnote{E-mail: nakayama@hep-th.phys.s.u-tokyo.ac.jp}}\\ 
\vspace{1cm}
\noindent{\small{\textit{Department of Physics, Faculty of Science, University of 
Tokyo}} \\ \vspace{2mm}
\small{\textit{Hongo 7-3-1, Bunkyo-ku, Tokyo 113-0033, Japan}}}
\end{center}
\vspace{1cm}
\begin{abstract}
We compute the index for the conifold gauge theory from type IIB supergravity (superstring) on $AdS_5 \times T^{1,1}$. We discuss its implication from the gauge theory viewpoint.

\end{abstract}

\end{titlepage}


\sectiono{Introduction}\label{sec:1}
Given a tremendous diversity of the ``landscape" of quantum field theories, the classification of conformal field theories (CFTs) is important because they are located at the special points of the ``landscape":  they are (attractive) IR fixed points a la Wilson. A key element of such a classification is invariants of CFTs under marginal deformations. In \cite{Kinney:2005ej}, it was argued that the index for the  four-dimensional superconformal field theories (SCFTs) on $S^3\times R$ plays such a role. Moreover it was discussed that all the information of these invariants are encoded\footnote{The other invariant of the SCFT under marginal deformations may be the central charge $a$. We guess there should be some connections.} in the (twisted) Witten index on $S^3 \times R$ \cite{Romelsberger:2005eg,Kinney:2005ej}
\begin{eqnarray}
\mathcal{I}^W(t,y) = \mathrm{Tr} (-1)^F e^{-\beta \Delta} t^{2(E+j_2)}y^{2j_1} \ , \label{Witten}
\end{eqnarray}
where $E$ is the energy (or conformal dimension via conformal mapping from $S^3 \times R$ to $R^4$) and $j_1, j_2$ denote the spin quantum number corresponding to the rotation around $S^3$ whose isometry is $SU(2)_1\times SU(2)_2$.  The regularizing factor $\Delta$ is given by the anti-commutator of a specific supercharge\footnote{We use the notation of $\dagger$ as the BPZ conjugation in the radial quantization scheme: if we go back to $R^4$ by conformal mapping, the BPZ conjugation of the SUSY transformation is given by the superconformal transformation: $\bar{Q}_{\dot{\alpha}}^\dagger = \bar{S}^{\dot{\alpha}}$.}
\begin{eqnarray}
\Delta = 2\{\bar{Q}^{-\frac{1}{2} \dagger}, \bar{Q}^{-\frac{1}{2}}\} = E-2j_2 - \frac{3}{2}r\ ,
\end{eqnarray}
which should be positive definite from the unitarity. Here we denote $r$ as the $U(1)$ R symmetry of the $\mathcal{N}=1 $ superconformal algebra $SU(2,2|1)$ \cite{Flato:1983te,Dobrev:1985qv} (see appendix \ref{sec:0}).

The index does not depend on any continuous parameters of the theory and is invariant under any (marginal) deformations of the SCFT unless it breaks the superconformal invariance. The only contribution to the index comes from the states with $\Delta = 0$ due to the Bose-Fermi cancellation. As an immediate consequence, the index does not depend on $\beta$.

By maximally using the fact that the index is independent of the coupling constant, the computation of the index for $\cN=4$ SYM theory and its orbifold descendants has been done in \cite{Kinney:2005ej,Nakayama:2005mf}. These theories are connected to free gauge theories by a marginal deformation, so the index can be computed in the free gauge theory limit, which simplifies the practical computation. 

It is interesting to compute the index for other SCFTs which are not connected to a free gauge theory by a marginal deformation. In this paper, we will compute the index for a gauge theory living on $N$ D3-branes placed at the conifold singularity (abbreviated as the ``conifold gauge theory") \cite{Klebanov:1998hh}. The theory has a conformal fixed point with nonzero anomalous dimension of chiral operators. Since the theory is intrinsically strongly coupled, the computation of the index seems hard.

This gauge theory has a dual gravity description as type IIB superstring theory on $AdS_5 \times T^{1,1}$ \cite{Klebanov:1998hh}. Thus we can compute the index for strongly coupled gauge theory from the AdS-CFT correspondence \cite{Aharony:1999ti} at least in the large $N$ limit by taking weakly coupled gravity limit. In this paper, we first compute the index for the conifold gauge theory from the Kaluza-Klein (KK) reduction of type IIB supergravity on  $AdS^5 \times T^{1,1}$. The KK reduction was thoroughly studied in \cite{Ceresole:1999zs,Ceresole:1999ht} (see also \cite{Gubser:1998vd,Jatkar:1999zk}), and we will use their results to compute the index. Our result is the first nontrivial example of the index for strongly coupled gauge theories that are not connected to free gauge theories by a marginal deformation. We also compare the result with the gauge theory side by showing operator correspondence discussed in \cite{Ceresole:1999zs}.

The organization of the paper is as follows. In section 2, we compute the index for the conifold gauge theory from the dual $AdS_5 \times T^{1,1}$ description. In section 3, we interpret our results from the gauge theory side. In section 4, we present the summary and related discussions. In appendices, we summarize several technical facts used in the main text. In appendix A, we present the superconformal algebra $SU(2,2|1)$. In appendix B, we summarize relevant unitary representations of $SU(2,2|1)$. In appendix C, we summarize the KK reduction of type IIB supergravity on $AdS_5 \times T^{1,1}$.

\sectiono{Index for Supergravity on $AdS_5 \times T^{1,1}$}\label{sec:2}
In this paper, we study the index for the superconformal field theory living on $N$ D3-branes placed at the conifold singularity that is defined as a curve 
\begin{eqnarray}
z_1^2 +z_2^2 + z_3^2 + z_4^2 = 0 \ ,
\end{eqnarray}
in $\mathbf{C}^4$. The singular point where we put D-branes is $z_1 = z_2 = z_3 = z_4 = 0$. The conifold admits a Calabi-Yau cone metric
\begin{eqnarray}
dS^2_{con} = dr^2 + r^2 ds^2_{T^{1,1}} \ ,
\end{eqnarray}
where $ds^2_{T^{1,1}}$ is the Sasaki-Einstein metric for the homogeneous coset space called $T^{1,1}$
\begin{eqnarray}
\frac{SU(2)\times SU(2)}{U(1)} \ .
\end{eqnarray}
If we take the near horizon geometry of the $N$ D3-brane placed at the conifold singularity, we obtain $AdS_5\times T^{1,1}$ geometry with $N$ units of background flux.\footnote{In the following, we always assume large $N$ limit.} According to the AdS-CFT correspondence, the superconformal field theory realized on these D3-branes is dual to type IIB supergravity on $AdS_5\times T^{1,1}$.

We would like to compute the index for the conifold gauge theory from the KK reduction of type IIB supergravity on $AdS_5\times T^{1,1}$. Our motivation to begin with the gravity side is that the dual gauge theory is not connected to a free gauge theory under marginal deformations unlike the case discussed in \cite{Kinney:2005ej,Nakayama:2005mf}. As a consequence the gauge theory is intrinsically strongly coupled and difficult to study directly. In the supergravity side, on the other hand, we can take the weakly coupled limit, which enables us to compute the index in a straightforward manner. 

It should be noted, however, that the index computed from the supergravity in general cannot be free from information of the dual gauge theory for the following reasons
\begin{itemize}
	\item Given a mass of the KK-mode, there is an ambiguity to assign AdS energy $E$ \cite{Klebanov:1999tb}. This happens e.g. for scalar with mass in the range
\begin{eqnarray}
-4 < m^2 <-3 \ .
\end{eqnarray}
There is no canonical way to choose one particular branch.\footnote{If we demand the conventional realization of superconformal algebra, they are uniquely determined in some cases.}
	\item Inclusion (or exclusion) of the singleton/doubleton representation should be fixed by hand. This, in part, corresponds to a freedom to include (or exclude) decoupled $U(1)$ degrees of freedom of the gauge theory. We will return to this problem in section \ref{sec:2}.
\end{itemize}

With these remarks in mind, let us compute the index 
\begin{eqnarray}
\mathcal{I}^W(t,y) = \mathrm{Tr} (-1)^F e^{-\beta \Delta} t^{2(E+j_2)}y^{2j_1} \ , \label{Witten}
\end{eqnarray}
for the conifold gauge theory. We first begin with the single particle index
\begin{eqnarray}
\mathcal{I}_{sp}^W(t,y) = \mathrm{Tr}_{sp} (-1)^F e^{-\beta \Delta} t^{2(E+j_2)}y^{2j_1} \ , 
\end{eqnarray}
where $\mathrm{Tr}_{sp}$ is taken over all single particle KK states on $AdS_5\times T^{1,1}$. The direct KK reduction of type IIB supergravity on $AdS_5\times T^{1,1}$ was almost completed in \cite{Ceresole:1999zs,Ceresole:1999ht}, where some multiplets (involving vector-spinor harmonics) are correctly guessed by filling the multiplets from the superconformal algebra. 

The KK modes are labeled by three quantum numbers $(s,l,r)$ corresponding to the isometry of $T^{1,1}$ $SU(2)\times SU(2) \times U(1)$. We will identify this $U(1)$ symmetry as the R-symmetry of the superconformal algebra. For general values of $(s,l,r)$, the KK states are in the long multiplet and do not contribute to the index. As discussed in the introduction, the only contribution to the index comes from $\Delta = E - \frac{3}{2}r - 2j_2 = 0$ states. They are either LH-chiral multiplets or LH-semi-long multiplets as reviewed in appendix \ref{sec:A}. We summarize the KK reduction of type IIB supergravity on $AdS_5\times T^{1,1}$ and the BPS states contributing to the index in appendix \ref{sec:B}.

Collecting all the BPS states contributing to the index from table \ref{tableA}-\ref{tableI} {\it except for Betti multiplets}, we have 
\begin{align}
\mathcal{I}_{sp}^W(t,y) &= \frac{1}{(1-yt^3)(1-y^{-1}t^3)}\left(-1+t^6 + \sum_{r=0}^\infty -(r+1)^2 t^{3r+9} (y+y^{-1}) + \sum_{r=1}^\infty - r^2 t^{3r} (y+y^{-1}) \right.  \cr &+ \sum_{r=1}^\infty 2r(r+2) t^{3r+6} (y+y^{-1})  + \sum_{r=-1}^\infty (r+2)^2 t^{3r+9} + \sum_{r=1}^{\infty} r^2t^{3r+9} + \sum_{r=0}^\infty (r+1)^2 t^{3r}  \cr &+ \left. \sum_{r=0}^\infty - 2(r+1)(r+3)t^{3r+6} + \sum_{r=2}^{\infty} (r-1)^2 t^{3r} + \sum_{r=2}^\infty -2(r-1)(r+1)t^{3r+6} \right) \cr
&= \frac{4t^3}{1-t^3} -\frac{yt^3}{1-yt^3} - \frac{y^{-1}t^3}{1-y^{-1}t^3} \ ,
\label{sp}
\end{align}
where we have included some singleton representations as was done in \cite{Kinney:2005ej,Nakayama:2005mf}. We need some explanations of several factors here. First of all $\frac{1}{(1-yt^3)(1-y^{-1}t^{3})}$ comes from the $SU(2,2)$ descendants obtained by acting $P_{-+}$ and $P_{++}$ to the BPS saturating $SU(2,2)$ primary states $\Delta_0 = E_0 - \frac{3}{2}r_0 - 2j_{20} = 0$. $-1$ in the first line of the parenthesis is to subtract the contribution from the identity operator in the gauge theory side which appears in $k=0$ Vector Multiplet III.\footnote{However, one should note that the KK spectrum corresponding to the state does not vanish as we will discuss later. Instead it gives a so-called Betti multiplet by taking other branch of AdS energy.} The second $t^6$ in the parenthesis comes from the fact that the $SU(2,2)$ descendants for Gravitino Multiplet I at $k=0$ has a BPS saturated null vector at level 1 (corresponding to the Dirac equation in the gauge theory side).
 
The final expression \eqref{sp} is remarkable and we can now exponentiate it to obtain multi-particle contributions to the index  as
\begin{eqnarray}
\mathcal{I}^W_{T^{1,1}}(t,y) = \exp\left[\sum_{n=1}^{\infty} \frac{1}{n} \mathcal{I}_{sp}(t^n,y^n) \right] = \prod_{n=1}^\infty \frac{(1-y^{-n}t^{3n})(1-y^nt^{3n})} { (1-t^{3n})^4} \ .
\end{eqnarray}

We now supplement additional states contributing to the index. Their existence is related to nontrivial homology of $T^{1,1}$ ($b^2 = b^3 = 1$), and they are called Betti multiplets. In our case, they are formally obtained from massless ($k=0$) Vector Multiplet I from RR 4-form with one $AdS_5$ index and three $T^{1,1}$ indices (as harmonic 3-form on $T^{1,1}$). It is singlet under $SU(2)\times SU(2) \times U(1)$ and identified with the conserved baryon current multiplet (see the next section for a gauge theory interpretation). There is another massless scalar from type III Vector Multiplet, where we naively expect semi-long multiplets but the KK expansion gives complementary chiral multiplets \cite{Ceresole:1999zs,Ceresole:1999ht}. The massless scalar is related to the 2-form wrapped on the non-trivial 2-cycle (as a harmonic form) of $T^{1,1}$, which yields the second modulus of the theory. It seems that these two contributions to the index cancel with each other.

We also note that in the above discussion, we have not included the contribution to the index from asymptotically $AdS_5$ BPS black hole excitations. It has been conjectured that these contributions to the index will cancel out in the large $N$ limit \cite{Kinney:2005ej} and indeed they do in the $N=4$ SYM theory and its orbifold descendants. We expect that similar things happen in our case as well. 

Before moving on to the gauge theory discussion, we would like to consider some generalizations of the index.
Combining it with the internal symmetry of the gauge theory under consideration, we can further twist the index to obtain more information about the SCFT. For instance,
the index for the conifold gauge theory can be further twisted by internal flavor symmetries $SU(2)\times SU(2)$ which commute with the superconformal algebra. Defining 
\begin{eqnarray}
\mathcal{I}^W_{T^{1,1}}(t,y,w,z) = \mathrm{Tr} (-1)^F e^{-\beta \Delta} t^{2(E+j_2)}y^{2j_1} w^{2s_3} z^{2l_3} \ ,
\end{eqnarray}
and repeating the computation of the index, we obtain
\begin{eqnarray}
\mathcal{I}^W_{T^{1,1}}(t,y,w,z) = \prod_{n=1}^\infty \frac{(1-y^{-n}t^{3n})(1-y^nt^{3n})} { (1-t^{3n}w^nz^n)(1-t^{3n}w^{-n}z^{-n})(1-t^{3n}w^nz^{-n})(1-t^{3n}w^{-n}z^n)} \ .
\end{eqnarray}

Yet another possible twist is the discrete symmetry discussed in \cite{Klebanov:1998hh}, which exists at particular points of the moduli space with zero background Kalb-Ramond $B$ field as we have implicitly assumed so far. We combine $z_4 \to -z_4$ symmetry with a worldsheet parity $\Omega (-1)^F$ so that it is non R-symmetry (and hence commutes with $\Delta$). In the gauge theory side, it corresponds to the exchange of two $SU(N)$ gauge group together with the charge conjugation $(N,\bar{N}) \to (\bar{N},N)$.\footnote{The combination of the two is necessary in order to keep the superpotential invariant.} The invariant single particle index is given by
\begin{eqnarray}
\mathcal{I}^W_{T^{1,1}:{sp/inv}}(t,y) = \frac{4t^3-5t^6+3t^9}{(1-t^3)^3} -\frac{yt^3}{1-yt^3} - \frac{y^{-1}t^3}{1-y^{-1}t^3} \ . \label{invs}
\end{eqnarray}
In this expression, the Betti multiplet is naturally projected out because they are odd under this discrete symmetry. In the last two terms, we find a single-particle index from the diagonal (decoupled) $U(1)$ vector multiplet as expected. The multi-particle index may also be computed from this expression.\footnote{We should note that the multi-particle contribution is {\it not} obtained by simply exponentiating \eqref{invs}. For instance, two-particle states from two single-particle states with odd parity should not be projected out.}

\sectiono{Toward the Gauge Theory Computation}\label{sec:2}
Now let us move on to the gauge theory interpretation of the index derived in the last section. The gauge theory living on $N$ D3-branes at the conifold singularity is described by $U(N)\times U(N)$ gauge theory coupled to two chiral superfields $A_i$ in the $(N,\bar{N})$ bifundamental representation and two chiral superfields $B_i$ in the $(\bar{N},N)$ bifundamental representation \cite{Klebanov:1998hh}. The global symmetry is $SU(2)_1\times SU(2)_2 \times U(1)_r$, under which $A_{i_1}$ and $B_{i_2}$ transform as a doublet of two $SU(2)$s. 

The theory has a (nonrenormalizable) superpotential
\begin{eqnarray}
W = \epsilon^{ij} \epsilon^{kl}\mathrm{Tr} A_iB_k A_j B_l \ , \label{supp}
\end{eqnarray}
and accordingly the chiral multiplets $A_i$ and $B_j$ have $1/2$ unit of R-charge that is identified with the superconformal R-charge at the IR fixed point.

Under the renormalization group flow, one combination of the $U(1)$ vector multiplet becomes IR free and the other diagonal $U(1)$ vector multiplet decouples from the beginning. In order to obtain correct AdS-CFT correspondence, we have to remove the IR free $U(1)$ vector multiplet from the gauge theory side.\footnote{The removal of the IR free $U(1)$ vector multiplet is crucial while the removal of the other decoupled $U(1)$ part is rather arbitrary. The former will appear as baryon number {\it global} symmetry. The latter degree of freedom is related to the singleton representation in the gravity side, and we will include it for a cosmetic appearance of the index as has been done in \cite{Kinney:2005ej,Nakayama:2005mf}.}

Under the state operator mapping, the index
\begin{eqnarray}
\mathcal{I}^W(t,y) = \mathrm{Tr} (-1)^F e^{-\beta \Delta} t^{2(E+j_2)}y^{2j_1} \ , \label{Witteng}
\end{eqnarray}
on $S^3\times R$ counts a class of BPS operators with $\Delta = 0$ inserted at the origin of $R^4$. To compute \eqref{Witteng}, we first begin with the single-trace operators in the gauge theory side, which correspond to single particle states in the gravity side.
Actually, in \cite{Ceresole:1999zs}, all the single-particle BPS states from the KK reduction on $AdS_5\times T^{1,1}$ are identified with the single-trace BPS operators in the gauge theory. Concentrating on the $SU(2,2|1)$ primary states, we have the following map:

\begin{itemize}
	\item $\mathrm{Tr}(W_{\alpha}e^V\bar{W}_{\dot{\alpha}} e^{-V}(AB)^k) $: semi-long multiplets from Graviton Multiplet
	\item $\mathrm{Tr}(W_{\alpha} (AB)^k)$: chiral multiplets from Gravitino Multiplet I
	\item $\mathrm{Tr}(e^V\bar{W}_{\dot{\alpha}}e^{-V}(AB)^k)$: semi-long multiplets from Gravitino Multiplet III
	\item  $\mathrm{Tr}(e^V\bar{W}_{\dot{\alpha}}e^{-V}W^2(AB)^k)$: semi-long multiplets from Gravitino Multiplet IV 
	\item $\mathrm{Tr}(AB)^k$: chiral multiplets from Vector Multiplet I 
	\item $\mathrm{Tr}(W^2(AB)^k)$: chiral multiplet from Vector Multiplet IV
\end{itemize}
by setting $l=s= \frac{r}{2}$, and
 \begin{itemize}
	\item $\mathrm{Tr}(W_{\alpha}(Ae^V\bar{A}e^{-V})(AB)^k)$: semi-long multiplets from Gravitino Multiplet I
	\item $\mathrm{Tr}(Ae^V\bar{A}e^{-V}(AB)^k)$: semi-long multiplets from Vector Multiplet I
	\item $\mathrm{Tr}(Ae^{V}\bar{A}e^{-V}W^2(AB)^k)$: semi-long multiplets from Vector Multiplet IV
\end{itemize}
by setting $l=s-1 = \frac{r}{2}$ or similar operators $(A\to B)$ by setting $l-1=s=\frac{r}{2}$. Here the gauge indices are contracted in a covariant way, and all the $SU(2) \times SU(2)$ indices of $A$ and $B$ are separately symmetrized to obtain the maximum $SU(2)$ spin.\footnote{This constraint comes from the F-term condition. For example, $\mathrm{Tr}(AB)^k$ is in $(k/2,k/2)$ representation and $\mathrm{Tr}(Ae^{V}\bar{A}e^{-V}(AB)^k)$ is in $(k/2+1,k/2)$ representation.} We also note that the above expressions of the operators are rather schematic ones that are valid only for tree level action,\footnote{We have also suppressed possible dependence on (super)derivatives. For example, the energy momentum tensor supercurrent contains not only $W_{\alpha}\bar{W}_{\dot{\alpha}}$, but also $D_{\alpha}\Phi \bar{D}_{\dot{\alpha}} \bar{\Phi}$ even at the tree level.} and we have to take into account operator mixing and nonperturbative corrections to the chiral ring relation in the full interacting theory.

In addition to these multiplets, we also have a baryon number (super)current which is a conserved multiplet
\begin{eqnarray}
\mathcal{B} = \mathrm{Tr}( Ae^{V}\bar{A}e^{-V}) - \mathrm{Tr}(Be^{V}\bar{B}e^{-V}) \ ,
\end{eqnarray}
where $SU(2)$ indices are contracted {\it antisymmetrically} so that $\mathcal{B}$ is $SU(2)\times SU(2)$ singlet. In the gravity side, it corresponds to the Betti multiplet (formally massless multiplets from $k=0$ Vector Multiplet I) \cite{Ceresole:1999zs}. All the perturbative states are not charged under this baryonic current. As a nonperturbative object, the wrapped D-brane around nontrivial 3-cycle in $T^{1,1}$ has the baryonic charge and is known as a giant graviton \cite{Gubser:1998fp}. The corresponding operator is a dibaryon operator
\begin{eqnarray}
\mathcal{D} = \epsilon^{\alpha_1\cdots\alpha_N}\epsilon_{\beta_1\cdots\beta_N}\prod_{i=1}^N A_{\alpha_i}^{\ \beta_i} \ ,
\end{eqnarray}
which has a conformal dimension $\frac{3N}{4}$ consistent with the energy of wrapped D-brane. However, these operators do not appear in the large $N$ limit, and as a consequence, we have not included them in the index computed in the last section.

Finally, we have yet another massless scalar from harmonic two-form $A_{ab}$ on $T^{1,1}$ as a consequence of nonzero Betti number of $T^{1,1}$. This corresponds to a exactly marginal deformation of the Lagrangian that breaks the symmetry under $g_1 \leftrightarrow g_2$. 

In this way, we have a gauge theory interpretation of all the single particle KK states of type IIB supergravity on $AdS_5\times T^{1,1}$ contributing to the index as single-trace operators. Then we can construct multi-trace operators from these single-trace operators and we expect that the index obtained in the last section is fully recovered (albeit operator mixing should be treated properly).

However, this complete operator correspondence suggests that even the BPS partition function 
\begin{eqnarray}
\mathcal{Z}(t,y) = \mathrm{Tr}_{\Delta=0} t^{2(E+j_2)}y^{2j_1} \ , \label{part}
\end{eqnarray}
gives the identical answer in the gauge theory side (``weakly coupled limit") and in the supergravity side (``strongly coupled limit"). From the superconformal symmetry alone, we cannot expect such an agreement because in principle additional pairs of BPS multiplets can appear/disappear as a decay/merge of long multiplets. For example, it was shown that the BPS partition function of $N=4$ SYM at zero coupling is different from the BPS partition function of type IIB supergravity on $AdS_5\times S^5$.

It was observed in \cite{Kinney:2005ej} that {\it weakly coupled} $1/8$ BPS partition function of the $N=4$ SYM theory is identical to the $1/8$ BPS partition function of type IIB supergravity on $AdS_5\times S^5$ although the zero coupling partition function gives a different result. This should be understood as a kind of dynamical protection of BPS states and it essentially means that the only relevant deformation of the BPS sector is the deformation of the chiral ring, and drastic structure changes such as a decay or merge of long multiplets only occur in the zero coupling limit.

The deformation of the chiral ring structure is summarized by the generalized Konishi anomaly equation \cite{Konishi:1983hf,Konishi:1985tu,Cachazo:2002ry}:
\begin{eqnarray}
\bar{D}^2 [\Phi_i\bar{\Phi}_i] = \Phi_i \frac{\partial W}{\partial \Phi_i} +\frac{1}{32\pi^2} \mathrm{Tr}_iW^\alpha W_\alpha 
\end{eqnarray}
in its simplest form. These relations are believed to be exact operator identities in the full quantum theory.\footnote{Perturbatively it was proved as a generalization of the Adler-Bardeen theorem. Possible nonperturbative corrections may be forbidden by the symmetry argument in most examples.}

There are several ways to read this equation in the context of chiral ring, but one interpretation of this equation is that the F-term equation $ \partial_i W = \bar{D}^2 \bar{\Phi}_i \sim 0$ (leading to the classical chiral ring) is modified by the quantum corrections. One should note that if the superpotential $W$ contains as many fields as we have in the action, the chiral ring of the classical gauge theory is modified but the number of the independent elements is the same. If, on the other hand, $W$ contains less field, it is possible that the generalized Konishi anomaly equation truncates independent elements of the chiral ring.

For instance, let us consider the SQCD with no superpotential. Then the classical chiral primary operator $\mathrm{Tr}W^{\alpha}W_{\alpha}$ now becomes a descendant field
because the Konishi anomaly equation reads
\begin{eqnarray}
\bar{D}^2 [Q\bar{Q}] = \frac{1}{32\pi^2} \mathrm{Tr} W^\alpha W_{\alpha} \ .
\end{eqnarray}
The fact that $\mathrm{Tr}W^\alpha W_{\alpha}$ is not a chiral primary operator can also be checked by a perturbative computation of the anomalous dimension revealing $E>3$ \cite{Anselmi:1996mq}.\footnote{This is one of the reasons why we have apparently different number of chiral primary operators in electric theory and magnetic theory in \cite{Romelsberger:2005eg}.} In the case of the conifold gauge theory, we do not encounter this problem as long as we are in a generic place of the moduli space.

Finally, we would like to comment on  the total moduli space of the exactly marginal deformation for the conifold gauge theory. From Leigh-Strassler like argument \cite{Leigh:1995ep}, we expect two (complex) dimensional moduli space of symmetry-preserving exactly marginal deformations, one of which corresponds to dilaton-axion expectation value in type IIB supergravity and the other corresponds to harmonic two-forms on $T^{1,1}$ \cite{Klebanov:1998hh}.\footnote{In addition, there are three excatly marginal but symmetry-breaking deformations \cite{Benvenuti:2005wi} originating from spin $s=l=1$ harmonics of the internal metric (six of which becomes marginally irrelevant). We would like to thank S.~Benvenuti for bringing our attention to \cite{Benvenuti:2005wi}.} Since the index does not change under the exactly marginal deformation, our results can also be regarded as the index for such deformed theories. The following particular limits are interesting:
\begin{itemize}
	\item We can turn off the superpotential term \eqref{supp}. In this limit, the BPS partition function clearly jumps as is the case with $N=4$ SYM theory in the zero coupling limit discussed above. The index, however, is expected to be unchanged.
	\item We can take the zero gauge coupling limit for one of the $SU(N)$ gauge groups. If we further turn off the superpotential term, we will obtain the index for $N=1$ SQCD with $N_f=2N_c$ flavors known as the Seiberg fixed points (plus noninteracting $SU(N)$ SYM theory).
\end{itemize}
Although these statements are a direct consequence of the general property of the index, they seem highly nontrivial and worthwhile checking independently.

\sectiono{Summary and Discussion}
In this paper, we computed the index for the conifold gauge theory from the KK reduction of type IIB supergravity on $AdS_5\times T^{1,1}$ with some hindsight from the dual gauge theory. As discussed in the introduction, the index is the unique quantity invariant under any marginal deformations of the theory, and it should capture the information of the structure of the SCFT in four-dimension. It would be, thus, interesting to find a way to extract some quantitative features of the SCFTs from the index. For a preliminary discussion on this direction, let us compare the indices for several different SCFTs so far computed.

The $N=4$ SYM theory with $U(N)$ gauge group has the index \cite{Kinney:2005ej}
\begin{eqnarray}
\mathcal{I}^W_{N=4;U(N)}(t,y) = \prod_{n=1}^{\infty}\frac{(1-y^{-n}t^{3n})(1-y^nt^{3n})} { (1-t^{2n})^3} \ .\end{eqnarray}
The $N=2$ $Z_k$ orbifold quiver gauge theory with $U(N)^k$ gauge groups has the index \cite{Nakayama:2005mf}
\begin{eqnarray}
\mathcal{I}^W_{N=2;U(N)}(t,y) = \prod_{n=1}^{\infty}\frac{(1-y^{-n}t^{3n})^k(1-y^nt^{3n})^k} { (1-t^{2nk})^2(1-t^{2n})^k} \ .
\end{eqnarray}
In our paper, we have shown that the index for the conifold gauge theory is given by 
\begin{eqnarray}
\mathcal{I}^W_{con}(t,y) = \prod_{n=1}^\infty \frac{(1-y^{-n}t^{3n})(1-y^nt^{3n})} { (1-t^{3n})^4} \ .
\end{eqnarray}
Note that the conifold gauge theory is obtained from $Z_2$ orbifold theory by adding mass terms to adjoint chiral superfields. These indices may be rewritten in a more concise and suggestive way by using the Dedekind eta function $\eta = t^{1/24}\prod_{n=1}^\infty (1-t^n) $.

From these expressions, we first observe that the $y$ dependence is similar and simple in all three cases.\footnote{If we decouple $N=1$ $U(1)$ vector multiplets, $y$ dependence completely vanishes in every case.} Furthermore, it is quite curious to know the reason why we can rewrite the whole expression in such a compact form. The appearance of the infinite product structure suggests as if there existed fundamental oscillator-like building blocks (with both statistics) to generate BPS operators, whose origin might be understood from the supergravity side.

One interesting question is the connection between the  central charge $a=c$ and the index. The two quantity has similar properties: they are invariant under any marginal deformations of the SCFT \cite{Barnes:2004jj}, and according to \cite{Kinney:2005ej}, such a quantity is essentially unique. We then expect a Cardy-like formula connecting these two. From the supergravity dual viewpoint, the central charge is identified with the inverse of the volume of the Sasaki-Einstein manifold \cite{Gubser:1998vd} which governs the asymptotic growth of the KK states that somehow must be reflected in the asymptotic form of the index.\footnote{We observe, however, a naive guess works {\it oppositely}. When $a$ becomes smaller, the volume increases so that we have {\it more} KK states.}

Another interesting issue is the $1/N \sim g_s$ correction to the index. The index does not depend on $\lambda \sim \alpha'$ but nontrivially depend on $N$. From the gauge theory side, the correction is obviously due to trace relations and baryonic operators, but from the gravity side, the correction seems nontrivial. They are nonperturbative $ \Delta \mathcal{I}^W = O(t^{N})$ contributions coming from the `stringy exclusion principle' and/or `giant gravitons' (wrapped D-branes). It would be interesting to study these points further in detail.

In the context of the systematic classification of SCFT from the index, especially in view of the AdS-CFT correspondence, a
generalization to more complicated Sasaki-Einstein manifold ($Y^{p,q}$ \cite{Gauntlett:2004yd}, $L^{p,q,r}$ \cite{Cvetic:2005ft,Cvetic:2005vk} etc) is an intriguing and challenging problem. The eigenvalue problem of the scalar Laplacian has been studied in \cite{Kihara:2005nt,Oota:2005mr} and chiral primary operators of the dual gauge theories are discussed in \cite{Benvenuti:2005cz,Benvenuti:2005ja} in comparison with the spectrum of massless particle in the dual geometry.

We hope to return to these interesting questions in near future.

\section*{Acknowledgements}
The author would like to thank C.~Romelsberger for fruitful discussions through e-mail correspondence.
This research is supported in part by JSPS Research Fellowships
for Young Scientists.

\appendix\sectiono{Superconformal algebra $SU(2,2|1)$}\label{sec:0}
Relevant (anti-)commutation relations of the superconformal algebra $SU(2,2|1)$  for our discussion are
\begin{align}
[(J_2)^{\dot{\alpha}}_{\dot{\beta}}, (J_2)^{\dot{\gamma}}_{\dot{\delta}}] &= \delta^{\dot{\gamma}}_{\dot{\beta}} (J_2)^{\dot{\alpha}}_{\dot{\delta}} - \delta^{\dot{\alpha}}_{\dot{\delta}} (J_2)^{\dot{\gamma}}_{\dot{\beta}}   \ ,\cr
[H,P^{\alpha\dot{\beta}}] &= P^{\alpha\dot{\beta}}  \ ,\cr
[H,\bar{Q}^{\dot{\gamma}}] &= \frac{1}{2} \bar{Q}^{\dot{\gamma}} \ ,\cr
[r,\bar{Q}^{\dot{\gamma}}] &= \bar{Q}^{\dot{\gamma}}  \ ,\cr
[r,P^{\alpha\dot{\beta}}] &= 0 \ ,\cr
\{ \bar{S}_{\dot{\alpha}},\bar{Q}^{\dot{\beta}} \} &= (J_2)^{\dot{\beta}}_{\dot{\alpha}} +\delta^{\dot{\beta}}_{\dot{\alpha}} \left(\frac{H}{2} - \frac{3}{4} r\right)  \ ,
\end{align}
where upper dotted spinor indices denote $SU(2)_2$ fundamental representation and un-dotted spinor indices denote $SU(2)_1$ fundamental representation.

\sectiono{Representation of $SU(2,2|1)$}\label{sec:A}
In this appendix, we summarize the classification of unitary representations of the superconformal algebra $SU(2,2|1)$ \cite{Flato:1983te} (see appendix B of \cite{Freedman:1999gp} for a review). The representation of the superconformal algebra $SU(2,2|1)$ is naturally decomposed into a highest weight representation of its bosonic subalgebra $SU(2,2) \simeq SO(4,2)$ (conformal algebra) and $U(1)_r$ (R-symmetry). Accordingly, the highest weight representation of $SU(2,2|1)$ is classified by four quantum numbers $D(E_0,j_1,j_2;r)$. 

The unitary representation of $SU(2,2|1)$ with generic quantum number $E_0,j_1,j_2,r$ consists of a long multiplet with $16$ $SU(2,2)$ highest weight representation. However, if one of the following combination of the quantum number vanishes, the representation becomes short.
\begin{align}
n_1^{(+)} &= N(E_0 + 2j_1 + \frac{3}{2}r)  \cr
n_1^{(-)} &= N(E_0 - 2j_1 + \frac{3}{2}r-2)  \cr
n_2^{(+)} &= N(E_0 + 2j_2 - \frac{3}{2}r)  \cr
n_2^{(-)} &= N(E_0 - 2j_2 - \frac{3}{2}r-2)  \cr
s_1 &= N(j_1) ,   \ \ \mathrm{and} \ \ \ \ s_2=N(j_2) \ , \label{short}
\end{align}
where, following \cite{Freedman:1999gp}, we have introduced the false-function: $N(x) =1$ for $x\neq 0$ and $N(0) =0$.
 The multiplet shortening structure is summarized in table \ref{table1}, which is borrowed from \cite{Freedman:1999gp} with corrections.

What is relevant for computing the index for SCFT on $S^3 \times R$ is which (short) multiplet contributes to the index. Recalling that the states contributing  to the index  satisfies $\Delta = E - 2j_2 -\frac{2}{3} r = 0$, we find {\it chiral LH-multiplets} with quantum number $D(E_0,j,0;r)$ with $r=\frac{2}{3}E_0$ and {\it LH-semi-long multiplet} $D(E_0,j_1,j_2;r)$ with $r=\frac{2}{3}(E_0-2j_2-2)$ (together with their specific $SU(2,2)$ descendants) will contribute to the index. The structure of these short multiplets contributing to the index are presented in table \ref{table2} and \ref{table3}.

Let us now consider the contribution to the index from $SU(2,2)$ descendants obtained by acting $P_{\alpha\dot{\alpha}}$ to $SU(2,2)$ primaries. It is clear only $P_{\pm+}$ will generate descendants contributing to the index by acting on $SU(2,2)$ primaries with vanishing $\Delta$. However, there is a possibility that $SU(2,2)$ descendants show  a multiplet shortening \cite{Mack:1975je}. In this case, we have to subtract contributions from their null vectors. This happens when 
\begin{align}
(a) \ \ \ \ \ \ j_1j_2 &\neq 0 \ ,  \ \ \ E_0 = 2 + j_1 + j_2  \cr
(b) \ \  \ \ \ \ j_1j_2 &= 0    \ , \ \ \ E_0 = 1+j \cr
(c) \ \ j_1=j_2 &= 0 \ , \ \ \ E_0 = 0 \ .
\end{align}
In the main text, we have encountered the condition $(b)$ interpreted as the Dirac equation of the gaugino. 

\begin{table}[bp]
\begin{center}
\begin{tabular}{c|c|c} 
 Level & $SU(2,2)$ representation & Multiplicity \\\hline
 $0$ & $D(E_0,j_1,j_2;r)$ & 1  \\\hline
 $1$&$D(E_0+\frac{1}{2},j_1+\frac{1}{2},j_2;r-1)$  & $n_1^{(+)}$ \\
 & $D(E_0+\frac{1}{2},j_1-\frac{1}{2},j_2;r-1)$ &  $s_1n_1^{(-)}$\\
 & $D(E_0+\frac{1}{2},j_1,j_2-\frac{1}{2};r+1)$ & $s_2n_2^{(-)}$ \\
 & $D(E_0+\frac{1}{2},j_1,j_2+\frac{1}{2};r+1)$ & $n_2^{(+)}$ \\\hline
 $2$ & $D(E_0+1,j_1,j_2;r-2)$   & $n_1^{(+)}n_1^{(-)}$   \\
 &$D(E_0+1,j_1+\frac{1}{2},j_2+\frac{1}{2};r)$  &$n_1^{(+)}n_2^{(+)}$  \\
 &$D(E_0+1,j_1+\frac{1}{2},j_2-\frac{1}{2};r)$  &$s_2n_1^{(+)}n_2^{(-)}$ \\
 &$D(E_0+1,j_1-\frac{1}{2},j_2+\frac{1}{2};r)$  &$s_1n_1^{(-)}n_2^{(+)}$  \\
 & $D(E_0+1,j_1-\frac{1}{2},j_2-\frac{1}{2};r)$ &$s_1s_2n_1^{(-)}n_2^{(-)}$   \\
 & $D(E_0+1,j_1,j_2;r+2)$ &$ n_2^{(-)}n_2^{(+)}$ \\\hline
$3$ &$D(E_0+\frac{3}{2},j_1,j_2+\frac{1}{2};r-1)$  & $n_1^{(+)} n_1^{(-)} n_2^{(+)}$ \\
 & $D(E_0+\frac{3}{2},j_1,j_2-\frac{1}{2};r-1)$ & $s_2n_1^{(+)}n_1^{(-)}n_2^{(-)}$ \\
 & $D(E_0+\frac{3}{2},j_1-\frac{1}{2},j_2;r+1)$ &$s_1n_1^{(-)}n_2^{(+)}n_2^{(-)}$   \\
 & $D(E_0+\frac{3}{2},j_1+\frac{1}{2},j_2;r+1)$ & $n_1^{(+)}n_2^{(+)}n_2^{(-)}$  \\\hline
$4$ & $D(E_0+2,j_1,j_2;r)$ &$n_1^{(+)}n_1^{(-)} n_2^{(+)}n_2^{(-)}$   \\\hline
\end{tabular} 
\end{center}
\caption{A long multiplet of $SU(2,2|1)$ contains $16$ highest weight representations of $SU(2,2)$. When the unitarity bound is saturated or $j_1j_2=0$, multiplet shortening occurs as shown in the table.}
\label{table1}
\end{table}

\begin{table}[bp]
\begin{center}
\begin{tabular}{c|ccc} 
$E\backslash R$ & $r$ & $r-1$ & $r-2$ \\\hline
$E_0$ & $\diamondsuit (j,0)$  &  &  \\
$E_0 +\frac{1}{2}$ &  & $(j+\frac{1}{2},0)\oplus (j-\frac{1}{2},0)$ &  \\
$E_0 +1$ &  &  & $(j,0)$ \\\hline
\end{tabular}
\end{center}
\caption{Chiral LH-multiplets $D(E_0,j,0;r)$ with $r=\frac{2}{3}E_0$. The representation with $\diamondsuit$ (in the top component of the table) contributes to the index. When $j=0$, further shortening occurs.}
\label{table2}
\end{table}

\begin{table}[bp]
\begin{center}
\begin{tabular}{c|cccc} 
 $E\backslash R$ & $r+1$ & $r$ & $r-1$& $r-2$\\\hline
 $E_0$&  & $(j_1,j_2)$ & & \\
$E_0+\frac{1}{2}$ & $\diamondsuit (j_1,j_2+\frac{1}{2})$ &  & $(j_1+\frac{1}{2},j_2)$ & \\
 &  &  &$(j_1-\frac{1}{2},j_2)$ & \\
 $E_0+1$ &  & $(j_1+\frac{1}{2},j_2+\frac{1}{2})$ & &$(j_1,j_2)$ \\
&  &$(j_1-\frac{1}{2},j_2+\frac{1}{2})$  & & \\
$E_0+\frac{3}{2}$ &  &  &$(j_1,j_2+\frac{1}{2})$ & \\\hline
\end{tabular}
\end{center}
\caption{LH-semi-long multiplet $D(E_0,j_1,j_2;r)$ with $r=\frac{2}{3}(E_0-2j_2-2)$.} The representation with $\diamondsuit$ (in the level one of the table) contributes to the index. When $j_1 =0$, further shortening occurs.
\label{table3}
\end{table}

\newpage

\sectiono{KK reduction on $AdS_5\times T^{1,1}$}\label{sec:B}

In this appendix, we summarize the KK spectrum of type IIB supergravity on $AdS_5\times T^{1,1}$ \cite{Ceresole:1999zs,Ceresole:1999ht}. The eigenvalue of the  scalar Laplacian $H_0(s,l,r)$ on $T^{1,1}$ is given by
\begin{eqnarray}
H_{0}(s,l,r) = 6\left(s(s+1)+l(l+1) - \frac{r^2}{8}\right) \ , \label{sL}
\end{eqnarray}
where $s$ and $l$ denote the spin quantum numbers of $SU(2)\times SU(2)$ isometries of $T^{1,1}$ and $r \in \mathbf{Z}$ corresponds to the  $U(1)$ R symmetry of the $SU(2,2|1)$. $(s,l)$ takes both integers or both half-integers. The $U(1)$ R-charge  $r$ is restricted to the range of $|q+r| \le 2s$ and $|q-r| \le 2l$ with a certain integer $|q|\le 2$, depending on the spin of the KK field.

As observed in \cite{Ceresole:1999zs,Ceresole:1999ht}, all the KK spectra of type IIB supergravity have eigenvalues related to the scalar Laplacian \eqref{sL} with shifted arguments. To utilize this fact, we introduce the following notation
\begin{align}
H_{0}^{\pm}(s,l,r) &= H_0(s,l,r\pm 1) \cr
H_{0}^{\pm\pm}(s,l,r) &= H_0(s,l,r\pm 2) \ .
\end{align}

In the following tables (\ref{tableA}-\ref{tableI}), we present all the KK spectrum on $AdS_5\times T^{1,1}$ classified by the highest weight representation of $SU(2,2|1)$. In the tables, we attached labels $C$ (chiral) and $S$ (semi-long) that indicate possible multiplet shortening for a specific choice of $(s,l,r)$ (see the discussion below). Short representations with $\diamondsuit$ (and their $SU(2,2)$ descendants) will contribute to the index as reviewed in appendix \ref{sec:A}.

To obtain relevant multiplet shortenings, we need a specific choice of $(s,l,r)$ so that one of \ref{short} vanishes \cite{Ceresole:1999zs,Ceresole:1999ht}. This happens when\footnote{There is an additional possibility: $j=l=\frac{r-2}{2}$ for vector multiplet III. }
\begin{eqnarray}
s=l= \left|\frac{r}{2}\right| , \ \ \sqrt{H_0+4} = \frac{3}{2}|r| + 2 \ , \label{cond1}
\end{eqnarray}
or
\begin{eqnarray}
s=l-1= \left|\frac{r}{2}\right| , \ \ \sqrt{H_0+4} = \frac{3}{2}|r| + 4\ , \label{cond2}
\end{eqnarray}
or
\begin{eqnarray}
s-1=l= \left|\frac{r}{2}\right| , \ \ \sqrt{H_0+4} = \frac{3}{2}|r| + 4\ . \label{cond3}
\end{eqnarray} 

To see this, we first note that we demand $H_0 + 4$ to be a perfect square of a rational number in order to obtain any BPS saturated states contributing to the index. By setting $s = \frac{k}{2} + n_1$, $l=\frac{k}{2} + n_2$ and $r=k$  with $(n_1,n_2) \in \mathbf{Z}$, this means 
\begin{eqnarray}
[2(n_1+n_2+1) + \frac{3}{2}k]^2 + 2(n_1^2 + n_2^2 - 4n_1n_2 -n_1-n_2) \label{persq}
\end{eqnarray}
should be a perfect square of a certain rational number.\footnote{One obvious way to obtain KK states with rational AdS energy is to require that the last parenthesis in \eqref{persq} vanishes \cite{Gubser:1998vd}. The Diophantine problem 
\begin{eqnarray}
n_1^2 + n_2^2 - 4n_1n_2 -n_1-n_2 = 0
\end{eqnarray}
was solved by any consecutive numbers $\{0,0,1,5,20,76,285,...\}$ generated by
\begin{eqnarray}
(n_1,n_2) = (f_-(f_+^{(i)}(0)),f_+^{(i)}(0))
\end{eqnarray}
with 
\begin{eqnarray}
f_{\pm}(n) = \frac{1+4n\pm\sqrt{1+12n+12n^2}}{2} \ , \ \ \ f_+^{(i)}(x) = \overbrace{f_+(f_+(\cdots f_+(x)))}^{i\mathrm{-times}} \ .
\end{eqnarray} 
Of course there are infinitely many other possibilities to obtain rational AdS energy. E.g. $k=0,n_1=3, n_2=4$ gives one example not generated by the above series.}

We can further reduce the number of relevant states for our index computation by requiring that they satisfy the BPS condition $E = 2j_2+ \frac{3}{2}r$. Under the BPS condition, the left hand side of \eqref{persq} should be $(\frac{3}{2}k + c)^2$ with $|c|\le 4$. Accordingly, possible quantum number is restricted to \eqref{cond1},\eqref{cond2},\eqref{cond3}.

The states contributing to the index are 
\begin{itemize}
	\item $-(r+1)^2$ semi-long multiplets from Graviton Multiplet
	\item $-r^2$ chiral multiplets from Gravitino Multiplet I
	\item $(r+2)^2$ semi-long multiplets from Gravitino Multiplet III
	\item $r^2$ semi-long multiplets from Gravitino Multiplet IV 
	\item $(r+1)^2$ chiral multiplets from Vector Multiplet I 
	\item $(r-1)^2$ chiral multiplet from Vector Multiplet IV
\end{itemize}
by setting $s=l= \frac{r}{2}$, and
 \begin{itemize}
	\item $2r(r+2)$ semi-long multiplets from Gravitino Multiplet I
	\item $-2(r+1)(r+3)$ semi-long multiplets from Vector Multiplet I
	\item $-2(r-1)(r+1)$ semi-long multiplets from Vector Multiplet IV
\end{itemize}
by setting $s=l-1 = \frac{r}{2}$ or $s-1=l=\frac{r}{2}$, where the signature in front of the multiplicity denotes the statistics of the corresponding $SU(2,2)$ highest weight states that contribute to the index.

\begin{table}[bp]
\begin{center}
\begin{tabular}{|l|c|c|c|c|} 
\multicolumn{5}{c}{Graviton Multiplet}  \ \ $E_0 = 1+\sqrt{H_0+4}$ \\\hline
  & $(j_1,j_2)$ & $E$  & $R$ & field  \\\hline
 S  & $(1,1)$  & $E_0+1$ &  $r$ & $H_{\mu\nu}$  \\\hline
 S  &  $(1,1/2)$ & $E_0+1/2$  & $r-1$ & $\psi_\mu^L$   \\
 S$^\diamondsuit$  &  $(1/2,1)$ & $E_0+1/2$  & $r+1$ & $\psi_\mu^R$   \\
 S  &  $(1/2,1)$ & $E_0+3/2$  & $r-1$ & $\psi_\mu^R$   \\
   &  $(1,1/2)$ & $E_0+3/2$  & $r+1$ &  $\psi_\mu^L$  \\\hline
  S &  $(1/2,1/2)$ & $E_0$ & $r$ &  $\phi_\mu$  \\
   &  $(1/2,1/2)$ & $E_0+1$  & $r+2$ & $a_\mu$   \\
  S &  $(1/2,1/2)$ & $E_0+1$   & $r-2$ & $a_\mu$  \\
   & $(1/2,1/2)$  & $E_0+2$  & $r$ &  $B_\mu$  \\\hline
   & $(1,0)$ &  $E_0+1$  &$r$  & $b_{\mu\nu}^+$ \\
  S & $(0,1)$ &   $E_0+1$   & $r$ & $b_{\mu\nu}^-$  \\\hline
   & $(1/2,0)$  & $E_0+1/2$   & $r+1$ &  $\lambda_L$  \\
  S & $(0,1/2)$ & $E_0+1/2$  & $r-1$ &  $\lambda_R$  \\
   & $(1/2,0)$ & $E_0+3/2$   & $r-1$ &  $\lambda_L$  \\
   & $(0,1/2)$ & $E_0+3/2$  & $r+1$ &   $\lambda_R$ \\\hline
   & $(0,0)$  & $E_0+1$  & $r$ &   $B$ \\\hline
\end{tabular}
\end{center}
\caption{}
\label{tableA}
\end{table}

\begin{table}[bp]
\begin{center}
\begin{tabular}{|l|l|c|c|c|c|} 
\multicolumn{6}{c}{Gravitino Multiplet I}  \ \ $E_0 = -\frac{1}{2}+ \sqrt{H_0^{-} +4}$ \\\hline
  &  &$(j_1,j_2)$  & $E$ &$R$ & field   \\\hline
   & S & $(1,1/2)$  &$E_0+1$  & $r$  & $\psi^L_\mu$ \\\hline
   & S$^\diamondsuit$ & $(1/2,1/2)$  & $E_0+1/2$   & $r+1$ & $\phi_\mu$ \\
   & S & $(1/2,1/2)$  & $E_0+3/2$   & $r-1$  &  $a_\mu$ \\\hline
 C  & S & $(1,0)$  & $E_0+1/2$ & $r-1$ & $a_{\mu\nu}$ \\
   &  & $(1,0)$    & $E_0+3/2$ & $r+1$ & $b^+_{\mu\nu}$ \\\hline
 C$^\diamondsuit$  & S & $(1/2,0)$  &$E_0$  & $r$ &  $\psi_L^{(T)}$ \\
 C  & S & $(1/2,0)$ & $E_0+1$ & $r-2$ & $\psi_L^{(T)}$ \\
   & S &  $(0,1/2)$& $E_0+1$ & $r$ & $\lambda_R$ \\
   &  & $(1/2,0)$ & $E_0+1$  & $r+2$ &  $\psi_L^{(T)}$ \\
   &  &$(1/2,0)$  & $E_0+2$ & $r$ & $\psi_L^{(T)}$ \\\hline
 C  & S & $(0,0)$ & $E_0+1/2$ & $r-1$ & $a$ \\
   &  & $(0,0)$ &  $E_0+3/2$ & $r+1$ & $a$  \\\hline
\end{tabular}
\end{center}
\caption{}
\label{tableB}
\end{table}

\begin{table}[bp]
\begin{center}
\begin{tabular}{|c|c|c|c|} 
\multicolumn{4}{c}{Gravitino Multiplet II}  \ \ $E_0 = \frac{5}{2}+ \sqrt{H_0^{+} +4}$ \\\hline
$(j_1,j_2)$  & $E$ &$R$ & field   \\\hline
$(1,1/2)$  &$E_0+1$  & $r$  & $\psi^L_\mu$ \\\hline
 $(1/2,1/2)$  & $E_0+1/2$   & $r+1$ & $a_\mu$ \\
  $(1/2,1/2)$  & $E_0+3/2$   & $r-1$  &  $B_\mu$ \\\hline
$(1,0)$  & $E_0+1/2$ & $r-1$ & $b^+_{\mu\nu}$ \\
$(1,0)$    & $E_0+3/2$ & $r+1$ & $a_{\mu\nu}$ \\\hline
 $(1/2,0)$  &$E_0$  & $r$ &  $\psi_L^{(T)}$ \\
 $(1/2,0)$ & $E_0+1$ & $r-2$ & $\psi_L^{(T)}$ \\
  $(0,1/2)$& $E_0+1$ & $r$ & $\lambda_R$ \\
 $(1/2,0)$ & $E_0+1$  & $r+2$ &  $\psi_L^{(T)}$ \\
$(1/2,0)$  & $E_0+2$ & $r$ & $\psi_L^{(T)}$ \\\hline
$(0,0)$ & $E_0+1/2$ & $r-1$ & $a$ \\
$(0,0)$ &  $E_0+3/2$ & $r+1$ & $a$  \\\hline
\end{tabular}
\end{center}
\caption{}
\label{tableC}
\end{table}

\begin{table}[bp]
\begin{center}
\begin{tabular}{|l|c|c|c|c|} 
\multicolumn{5}{c}{Gravitino Multiplet III}  \ \ $E_0 = -\frac{1}{2}+\sqrt{H_0^++4}$ \\\hline
  & $(j_1,j_2)$ & $E$  & $R$ & field  \\\hline
 S  & $(1/2,1)$  & $E_0+1$ &  $r$ & $\psi_\mu^R$  \\\hline
 S  &  $(1/2,1/2)$ & $E_0+1/2$  & $r-1$ & $\phi_\mu$   \\
   &  $(1/2,1/2)$ & $E_0+3/2$  & $r+1$ & $a_\mu$   \\\hline
  S$^\diamondsuit$  &  $(0,1)$ & $E_0+1/2$  & $r+1$ & $a_{\mu\nu}$   \\
 S  &  $(0,1)$ & $E_0+3/2$  & $r-1$ &  $b^-_{\mu\nu}$  \\\hline
  S &  $(0,1/2)$ & $E_0$ & $r$ &  $\psi^{(T)}_R$  \\
   &  $(0,1/2)$ & $E_0+1$  & $r+2$ & $\psi^{(T)}_R$   \\
   &  $(1/2,0)$ & $E_0+1$   & $r$ & $\lambda_L$  \\
 S  & $(0,1/2)$  & $E_0+1$  & $r-2$ &  $\psi_R^{(T)}$  \\
   & $(0,1/2)$ &  $E_0+2$  &$r$  & $\psi_R^{(T)}$ \\\hline
   & $(0,0)$ &   $E_0+1/2$   & $r+1$ & $a$  \\
   & $(0,0)$  & $E_0+3/2$   & $r-1$ &  $a$  \\\hline
\end{tabular}
\end{center}
\caption{}
\label{tableD}
\end{table}

\begin{table}[bp]
\begin{center}
\begin{tabular}{|l|c|c|c|c|} 
\multicolumn{5}{c}{Gravitino Multiplet IV}  \ \ $E_0 =\frac{5}{2}+\sqrt{H_0^{-}+4}$ \\\hline
  & $(j_1,j_2)$ & $E$  & $R$ & field  \\\hline
 S  & $(1/2,1)$  & $E_0+1$ &  $r$ & $\psi_\mu^R$  \\\hline
 S  &  $(1/2,1/2)$ & $E_0+1/2$  & $r-1$ & $a_\mu$   \\
   &  $(1/2,1/2)$ & $E_0+3/2$  & $r+1$ & $B_\mu$   \\\hline
  S$^\diamondsuit$  &  $(0,1)$ & $E_0+1/2$  & $r+1$ & $b^-_{\mu\nu}$   \\
 S  &  $(0,1)$ & $E_0+3/2$  & $r-1$ &  $a_{\mu\nu}^-$  \\\hline
  S &  $(0,1/2)$ & $E_0$ & $r$ &  $\psi^{(T)}_R$  \\
   &  $(0,1/2)$ & $E_0+1$  & $r+2$ & $\psi^{(T)}_R$   \\
   &  $(1/2,0)$ & $E_0+1$   & $r$ & $\lambda_L$  \\
 S  & $(0,1/2)$  & $E_0+1$  & $r-2$ &  $\psi_R^{(T)}$  \\
   & $(0,1/2)$ &  $E_0+2$  &$r$  & $\psi_R^{(T)}$ \\\hline
   & $(0,0)$ &   $E_0+1/2$   & $r+1$ & $a$  \\
   & $(0,0)$  & $E_0+3/2$   & $r-1$ &  $a$  \\\hline
\end{tabular}
\end{center}
\caption{}
\label{tableE}
\end{table}

\begin{table}[bp]
\begin{center}
\begin{tabular}{|l|l|c|c|c|c|} 
\multicolumn{6}{c}{Vector Multiplet I}  \ \ $E_0 = -2+ \sqrt{H_0+4}$ \\\hline
  &  &$(j_1,j_2)$  & $E$ &$R$ & field   \\\hline
   & S & $(1/2,1/2)$  &$E_0+1$  & $r$  & $\phi_\mu$ \\\hline
 C  & S & $(1/2,0)$  & $E_0+1/2$   & $r-1$ & $\psi^{(L)}_L$ \\
   &  S$^\diamondsuit$ & $(0,1/2)$  & $E_0+1/2$   & $r+1$  &  $\psi_R^{(L)}$ \\
   & S & $(0,1/2)$  & $E_0+3/2$ & $r-1$ & $\psi_R^{(L)}$ \\
   &  & $(1/2,0)$    & $E_0+3/2$ & $r+1$ & $\psi_L^{(L)}$ \\\hline
C$^\diamondsuit$  & S & $(0,0)$  &$E_0$  & $r$ &  $b$ \\
 C  & S & $(0,0)$ & $E_0+1$ & $r-2$ & $\phi$ \\
   &  & $(0,0)$ & $E_0+1$ & $r+2$ & $\phi$ \\
   &  & $(0,0)$ &  $E_0+2$ & $r$ & $\phi$  \\\hline
\end{tabular}
\end{center}
\caption{}
\label{tableF}
\end{table}

\begin{table}[bp]
\begin{center}
\begin{tabular}{|c|c|c|c|} 
\multicolumn{4}{c}{Vector Multiplet II}  \ \ $E_0 = 4+ \sqrt{H_0+4}$ \\\hline
 $(j_1,j_2)$  & $E$ &$R$ & field   \\\hline
 $(1/2,1/2)$  &$E_0+1$  & $r$  & $B_\mu$ \\\hline
 $(1/2,0)$  & $E_0+1/2$   & $r-1$ & $\psi^{(L)}_L$ \\
 $(0,1/2)$  & $E_0+1/2$   & $r+1$  &  $\psi_R^{(L)}$ \\
 $(0,1/2)$  & $E_0+3/2$ & $r-1$ & $\psi_R^{(L)}$ \\
 $(1/2,0)$    & $E_0+3/2$ & $r+1$ & $\psi_L^{(L)}$ \\\hline
 $(0,0)$  &$E_0$  & $r$ &  $\phi$ \\
 $(0,0)$ & $E_0+1$ & $r-2$ & $\phi$ \\
 $(0,0)$ & $E_0+1$ & $r+2$ & $\phi$ \\
 $(0,0)$ &  $E_0+2$ & $r$ & $\pi$  \\\hline
\end{tabular}
\end{center}
\caption{}
\label{tableG}
\end{table}

\begin{table}[bp]
\begin{center}
\begin{tabular}{|c|c|c|c|c|} 
\multicolumn{5}{c}{Vector Multiplet III}  \ \ $E_0 = 1+ \sqrt{H_0^{++}+4}$ \\\hline
  &  $(j_1,j_2)$  & $E$ &$R$ & field   \\\hline
   & $(1/2,1/2)$  &$E_0+1$  & $r$  & $a_\mu$ \\\hline
  &  $(1/2,0)$  & $E_0+1/2$   & $r-1$ & $\psi^{(T)}_L$ \\
   & $(0,1/2)$  & $E_0+1/2$   & $r+1$  &  $\psi_R^{(T)}$ \\
   & $(0,1/2)$  & $E_0+3/2$ & $r-1$ & $\psi_R^{(T)}$ \\
 C  &$(1/2,0)$    & $E_0+3/2$ & $r+1$ & $\psi_L^{(T)}$ \\\hline
   & $(0,0)$  &$E_0$  & $r$ &  $a$ \\
   & $(0,0)$ & $E_0+1$ & $r-2$ & $\phi$ \\
C & $(0,0)$ & $E_0+1$ & $r+2$ & $\phi$ \\
 C & $(0,0)$ &  $E_0+2$ & $r$ & $a$  \\\hline
\end{tabular}
\end{center}
\caption{}
\label{tableH}
\end{table}

\begin{table}[bp]
\begin{center}
\begin{tabular}{|c|c|c|c|c|c|} 
\multicolumn{6}{c}{Vector Multiplet IV}  \ \ $E_0 = 1+ \sqrt{H_0^{--}+4}$ \\\hline
  &  &$(j_1,j_2)$  & $E$ &$R$ & field   \\\hline
   & S & $(1/2,1/2)$  &$E_0+1$  & $r$  & $a_\mu$ \\\hline
 C  & S & $(1/2,0)$  & $E_0+1/2$   & $r-1$ & $\psi^{(T)}_L$ \\
   &  S$^\diamondsuit$ & $(0,1/2)$  & $E_0+1/2$   & $r+1$  &  $\psi_R^{(T)}$ \\
   & S & $(0,1/2)$  & $E_0+3/2$ & $r-1$ & $\psi_R^{(T)}$ \\
   &  & $(1/2,0)$    & $E_0+3/2$ & $r+1$ & $\psi_L^{(T)}$ \\\hline
  C$^\diamondsuit$  & S & $(0,0)$  &$E_0$  & $r$ &  $a$ \\
 C  & S & $(0,0)$ & $E_0+1$ & $r-2$ & B \\
   &  & $(0,0)$ & $E_0+1$ & $r+2$ & $\phi$ \\
   &  & $(0,0)$ &  $E_0+2$ & $r$ & $a$  \\\hline
\end{tabular}
\end{center}
\caption{}
\label{tableI}
\end{table}

\newpage
\bibliographystyle{utcaps}
\bibliography{T11}

\providecommand{\href}[2]{#2}\begingroup\raggedright\begin{thebibliography}{10}

\bibitem{Kinney:2005ej}
J.~Kinney, J.~Maldacena, S.~Minwalla, and S.~Raju, ``An index for 4 dimensional
  super conformal theories,''
\href{http://www.arXiv.org/abs/hep-th/0510251}{{\tt hep-th/0510251}}.

\bibitem{Romelsberger:2005eg}
C.~Romelsberger, ``An index to count chiral primaries in N = 1 d = 4
  superconformal field theories,''
\href{http://www.arXiv.org/abs/hep-th/0510060}{{\tt hep-th/0510060}}.

\bibitem{Flato:1983te}
M.~Flato and C.~Fronsdal, ``REPRESENTATIONS OF CONFORMAL SUPERSYMMETRY,'' {\em
  Lett. Math. Phys.} {\bf 8} (1984)
159.

\bibitem{Dobrev:1985qv}
V.~K. Dobrev and V.~B. Petkova, ``ALL POSITIVE ENERGY UNITARY IRREDUCIBLE
  REPRESENTATIONS OF EXTENDED CONFORMAL SUPERSYMMETRY,'' {\em Phys. Lett.} {\bf
  B162} (1985)
127--132.

\bibitem{Nakayama:2005mf}
Y.~Nakayama, ``Index for orbifold quiver gauge theories,''
\href{http://www.arXiv.org/abs/hep-th/0512280}{{\tt hep-th/0512280}}.

\bibitem{Klebanov:1998hh}
I.~R. Klebanov and E.~Witten, ``Superconformal field theory on threebranes at a
  Calabi-Yau singularity,'' {\em Nucl. Phys.} {\bf B536} (1998) 199--218,
\href{http://www.arXiv.org/abs/hep-th/9807080}{{\tt hep-th/9807080}}.

\bibitem{Aharony:1999ti}
O.~Aharony, S.~S. Gubser, J.~M. Maldacena, H.~Ooguri, and Y.~Oz, ``Large N
  field theories, string theory and gravity,'' {\em Phys. Rept.} {\bf 323}
  (2000) 183--386,
\href{http://www.arXiv.org/abs/hep-th/9905111}{{\tt hep-th/9905111}}.

\bibitem{Ceresole:1999zs}
A.~Ceresole, G.~Dall'Agata, R.~D'Auria, and S.~Ferrara, ``Spectrum of type IIB
  supergravity on AdS(5) x T(11): Predictions on N = 1 SCFT's,'' {\em Phys.
  Rev.} {\bf D61} (2000) 066001,
\href{http://www.arXiv.org/abs/hep-th/9905226}{{\tt hep-th/9905226}}.

\bibitem{Ceresole:1999ht}
A.~Ceresole, G.~Dall'Agata, and R.~D'Auria, ``KK spectroscopy of type IIB
  supergravity on AdS(5) x T(11),'' {\em JHEP} {\bf 11} (1999) 009,
\href{http://www.arXiv.org/abs/hep-th/9907216}{{\tt hep-th/9907216}}.

\bibitem{Gubser:1998vd}
S.~S. Gubser, ``Einstein manifolds and conformal field theories,'' {\em Phys.
  Rev.} {\bf D59} (1999) 025006,
\href{http://www.arXiv.org/abs/hep-th/9807164}{{\tt hep-th/9807164}}.

\bibitem{Jatkar:1999zk}
D.~P. Jatkar and S.~Randjbar-Daemi, ``Type IIB string theory on AdS(5) x T(n
  n'),'' {\em Phys. Lett.} {\bf B460} (1999) 281--287,
\href{http://www.arXiv.org/abs/hep-th/9904187}{{\tt hep-th/9904187}}.

\bibitem{Klebanov:1999tb}
I.~R. Klebanov and E.~Witten, ``AdS/CFT correspondence and symmetry breaking,''
  {\em Nucl. Phys.} {\bf B556} (1999) 89--114,
\href{http://www.arXiv.org/abs/hep-th/9905104}{{\tt hep-th/9905104}}.

\bibitem{Gubser:1998fp}
S.~S. Gubser and I.~R. Klebanov, ``Baryons and domain walls in an N = 1
  superconformal gauge theory,'' {\em Phys. Rev.} {\bf D58} (1998) 125025,
\href{http://www.arXiv.org/abs/hep-th/9808075}{{\tt hep-th/9808075}}.

\bibitem{Konishi:1983hf}
K.~Konishi, ``ANOMALOUS SUPERSYMMETRY TRANSFORMATION OF SOME COMPOSITE
  OPERATORS IN SQCD,'' {\em Phys. Lett.} {\bf B135} (1984)
439.

\bibitem{Konishi:1985tu}
K.-i. Konishi and K.-i. Shizuya, ``FUNCTIONAL INTEGRAL APPROACH TO CHIRAL
  ANOMALIES IN SUPERSYMMETRIC GAUGE THEORIES,'' {\em Nuovo Cim.} {\bf A90}
  (1985)
111.

\bibitem{Cachazo:2002ry}
F.~Cachazo, M.~R. Douglas, N.~Seiberg, and E.~Witten, ``Chiral rings and
  anomalies in supersymmetric gauge theory,'' {\em JHEP} {\bf 12} (2002) 071,
\href{http://www.arXiv.org/abs/hep-th/0211170}{{\tt hep-th/0211170}}.

\bibitem{Anselmi:1996mq}
D.~Anselmi, M.~T. Grisaru, and A.~Johansen, ``A Critical Behaviour of Anomalous
  Currents, Electric- Magnetic Universality and CFT$_4$,'' {\em Nucl. Phys.}
  {\bf B491} (1997) 221--248,
\href{http://www.arXiv.org/abs/hep-th/9601023}{{\tt hep-th/9601023}}.

\bibitem{Leigh:1995ep}
R.~G. Leigh and M.~J. Strassler, ``Exactly marginal operators and duality in
  four-dimensional N=1 supersymmetric gauge theory,'' {\em Nucl. Phys.} {\bf
  B447} (1995) 95--136,
\href{http://www.arXiv.org/abs/hep-th/9503121}{{\tt hep-th/9503121}}.

\bibitem{Benvenuti:2005wi}
S.~Benvenuti and A.~Hanany, ``Conformal manifolds for the conifold and other
  toric field theories,'' {\em JHEP} {\bf 08} (2005) 024,
\href{http://www.arXiv.org/abs/hep-th/0502043}{{\tt hep-th/0502043}}.

\bibitem{Barnes:2004jj}
E.~Barnes, K.~Intriligator, B.~Wecht, and J.~Wright, ``Evidence for the
  strongest version of the 4d a-theorem, via a-maximization along RG flows,''
  {\em Nucl. Phys.} {\bf B702} (2004) 131--162,
\href{http://www.arXiv.org/abs/hep-th/0408156}{{\tt hep-th/0408156}}.

\bibitem{Gauntlett:2004yd}
J.~P. Gauntlett, D.~Martelli, J.~Sparks, and D.~Waldram, ``Sasaki-Einstein
  metrics on S(2) x S(3),'' {\em Adv. Theor. Math. Phys.} {\bf 8} (2004)
  711--734,
\href{http://www.arXiv.org/abs/hep-th/0403002}{{\tt hep-th/0403002}}.

\bibitem{Cvetic:2005ft}
M.~Cvetic, H.~Lu, D.~N. Page, and C.~N. Pope, ``New Einstein-Sasaki spaces in
  five and higher dimensions,'' {\em Phys. Rev. Lett.} {\bf 95} (2005) 071101,
\href{http://www.arXiv.org/abs/hep-th/0504225}{{\tt hep-th/0504225}}.

\bibitem{Cvetic:2005vk}
M.~Cvetic, H.~Lu, D.~N. Page, and C.~N. Pope, ``New Einstein-Sasaki and
  Einstein spaces from Kerr-de Sitter,''
\href{http://www.arXiv.org/abs/hep-th/0505223}{{\tt hep-th/0505223}}.

\bibitem{Kihara:2005nt}
H.~Kihara, M.~Sakaguchi, and Y.~Yasui, ``Scalar Laplacian on Sasaki-Einstein
  manifolds Y(p,q),'' {\em Phys. Lett.} {\bf B621} (2005) 288--294,
\href{http://www.arXiv.org/abs/hep-th/0505259}{{\tt hep-th/0505259}}.

\bibitem{Oota:2005mr}
T.~Oota and Y.~Yasui, ``Toric Sasaki-Einstein manifolds and Heun equations,''
\href{http://www.arXiv.org/abs/hep-th/0512124}{{\tt hep-th/0512124}}.

\bibitem{Benvenuti:2005cz}
S.~Benvenuti and M.~Kruczenski, ``Semiclassical strings in Sasaki-Einstein
  manifolds and long operators in N = 1 gauge theories,''
\href{http://www.arXiv.org/abs/hep-th/0505046}{{\tt hep-th/0505046}}.

\bibitem{Benvenuti:2005ja}
S.~Benvenuti and M.~Kruczenski, ``From Sasaki-Einstein spaces to quivers via
  BPS geodesics: L(p,q|r),''
\href{http://www.arXiv.org/abs/hep-th/0505206}{{\tt hep-th/0505206}}.

\bibitem{Freedman:1999gp}
D.~Z. Freedman, S.~S. Gubser, K.~Pilch, and N.~P. Warner, ``Renormalization
  group flows from holography supersymmetry and a c-theorem,'' {\em Adv. Theor.
  Math. Phys.} {\bf 3} (1999) 363--417,
\href{http://www.arXiv.org/abs/hep-th/9904017}{{\tt hep-th/9904017}}.

\bibitem{Mack:1975je}
G.~Mack, ``ALL UNITARY RAY REPRESENTATIONS OF THE CONFORMAL GROUP SU(2,2) WITH
  POSITIVE ENERGY,'' {\em Commun. Math. Phys.} {\bf 55} (1977)
1.

\end{thebibliography}\endgroup

\end{document}